\documentclass[aps,prx,twocolumn,longbibliography]{revtex4-1}
\usepackage[usenames,dvipsnames]{color}
\usepackage{graphicx}
\usepackage{dcolumn}
\usepackage{bm}
\usepackage{amsmath}
\usepackage{amsfonts}
\usepackage{psfrag}
\usepackage[bookmarks,colorlinks=false]{hyperref}
\usepackage{mathtools}
\usepackage{accents}

\usepackage[mathcal]{euscript}

\newcommand{\Eq}[1]{Eq.~(\ref{#1})}
\newcommand{\Eqs}[2]{Eqs.~(\ref{#1})-(\ref{#2})}

\begin{document}

\title{Optical Nonlocality in Polar Dielectrics}

\author{Christopher R. Gubbin}
\author{Simone De Liberato}
\email[Corresponding author: ]{s.de-liberato@soton.ac.uk}
\affiliation{School of Physics and Astronomy, University of Southampton, Southampton, SO17 1BJ, United Kingdom}

\begin{abstract}
Phonon polaritons localised in polar nanoresonators and superlattices are being actively investigated as promising platforms for mid-infrared nanophotonics.
Here we show that the nonlocal nature of the phonon response can strongly modify their nanoscale physics.
Using a nonlocal dielectric approach we study dielectric nanospheres and thin dielectric films taking into account optical phonons dispersion. We discover a rich nonlocal 
phenomenology, qualitatively different from the one of plasmonic systems. Our theory allows us to explain the recently reported discrepancy between theory and experiments in atomic-scale superlattices, and it provides a practical tool for the design of phonon-polariton nanodevices.
\end{abstract}
\maketitle

\section{Introduction}
Nanophotonics is predicated on the ability to concentrate and control light on length scales substantially below the diffraction limit \cite{Ballarini2019}. The necessary deep sub-wavelength self-sustaining electromagnetic oscillations are possible through cycling electromagnetic energy into the kinetic energy of charged particles \cite{Khurgin2017} as in the ever-growing field of plasmonics \cite{Schuller2010, Brongersma2015}. Plasmonic concentrators, reliant on hybridisation between photons and a free electron gas, are inefficient for operating frequencies out of resonance with the gases plasma frequency which is typically in the near ultra-violet. A promising alternative in the mid-infrared are surface phonon polariton (SPhP) light concentrators, which permit deep sub-wavelength light confinement by storing electromagnetic energy as vibrations of the crystal lattice \cite{Greffet2002,Hillenbrand2002,Caldwell2015a}. Localization of SPhPs in user defined resonators \cite{Caldwell2013} allows for extremely small mode volumes \cite{Ellis2016,Gubbin2017} in highly tuneable low-loss modes \cite{Spann2016,Gubbin2016,Dunkelberger2018,Passler2018}, with applications in sensing \cite{Berte2018}, nonlinear optics \cite{Gubbin2017b, Razdolski2018}, and the creation of nanophotonic circuitry \cite{Li2016,Li2018,Chaudhary2019}.\\
Typical electromagnetic theories are parameterised by frequency dependant dielectric functions in which a spatially local relationship between electric field and polarization is implicit. This approximation is known to fail in plasmonic systems of nanometric length scale \cite{Raza2011,Fernandez-Dominguez2012}, when longitudinal plasma waves induced in the electron gas by the transverse photon field at the particle boundary become significant. Ignoring this nonlocal effect leads to erroneous predictions of modal frequencies and significant overestimation of the achievable field confinement \cite{Ciraci2012,Mortensen2014,Wubs2017,Maack2017}. \\
Polar dielectrics support longitudinal optical (LO) phonons, analogous to longitudinal plasma waves in noble metals. As SPhP concentrators grow smaller the dielectric local theory used to model them is also expected to break down, as LO phonons hybridise with the photon field and perturb the system response. Far from being a nuisance, hybridisation of LO phonons and SPhPs has been suggested as a possible path toward electrical injection of SPhPs, and was recently demonstrated by using elongated silicon carbide (SiC) polytypes \cite{Gubbin2019}, where a weak phonon mode \cite{Paarmann2016} naturally crosses the SPhP dispersion.\\
In this article we develop a macroscopic nonlocal theory describing the optical response of polar crystals. Initially we use it to study 3C-SiC nanospheres discovering rich nonlocal phenomenology. We then study epsilon-near-zero (ENZ) resonances in AlN thin films and build on these results to model crystal hybrids comprised of aluminium nitride (AlN) / gallium nitride (GaN) atomic-scale superlattices. {Our method provides an agile, numerically lightweight toolset to simulate and understand the nonlocal response of nanoscale polar structures, without the drawbacks of numerically taxing molecular dynamics \cite{Chalopin2012} or density functional calculations \cite{Paudel2009, Ratchford2019}. By comparison to experimental data we demonstrate our approaches validity for a variety of systems down to the nanometre scale.}  Our theory allows us to explain recently reported discrepancies between predictions of the dielectric theory and the experimental data \cite{Ratchford2019}, providing the first experimental evidence of nonlocal effects in polar dielectrics and demonstrating its relevance as a design tool for SPhP devices. For the sake of clarity most of the technical details of calculations have been placed into the Appendices.\\

\begin{figure*}
    \includegraphics[width=0.8\textwidth]{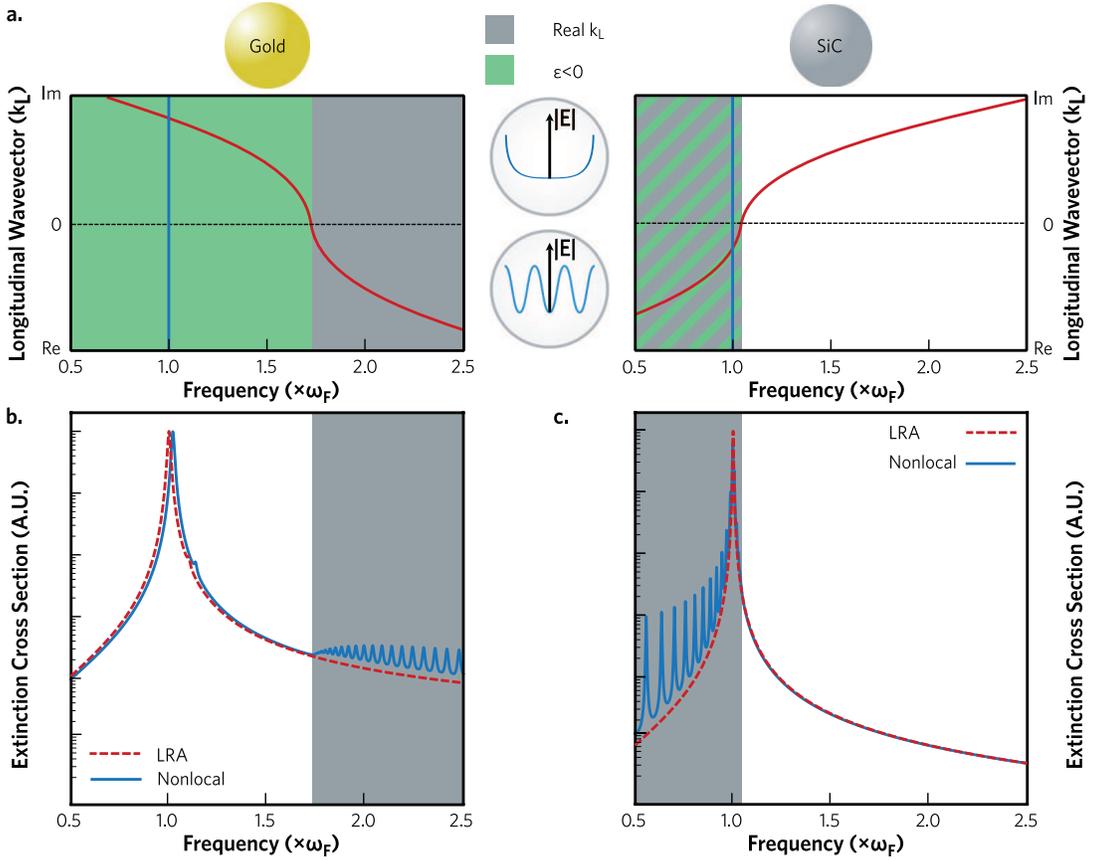}
	\caption{\label{fig:Fig1} a) Illustration of nonlocal effects in gold and SiC nanospheres. The two panels illustrate the dispersion of the longitudinal plasma (left) and  longitudinal optical phonon (right) as a function of frequency. Regions with real wavevector solutions are shaded grey. Green shading illustrates instead the region where the dielectric function is negative and can sustain surface modes, plasmonic or phononic.  The Fr{\"o}hlich resonance frequency, where $\epsilon_{\mathrm{LRA}} \left(\omega_{\mathrm{F}}\right)= - 2$, is shown by the solid blue line and the frequency axis is normalised to $\omega_{\mathrm{F}}$ in each case. The two spheres in the centre sketch the longitudinal mode electric field distribution in the case of imaginary (top) and real (bottom) wavevectors, relevant respectively for the metallic and dielectric cases. b,c) Extinction cross-section for a 10 nm radius b) gold and c) 3C-SiC nanosphere. Results calculated with a local response approximation (LRA) are shown by red-dashed curves and those calculated with a full nonlocal model by blue solid curves. {Parameters used in the calculations for both gold and 3C-SiC are reported in Appendix \ref{APC}}}.
\end{figure*}

\section{Nonlocal dielectric response}
The optical response of polar dielectrics can be described in the local response approximation (LRA) by assuming the electric $\mathbf{E}\left(\mathbf{r}, t\right)$ and displacement $\mathbf{D}\left(\mathbf{r}, t\right)$ fields are spatially local
\begin{equation}
	\mathbf{D}\left(\mathbf{r}, \omega\right) = \epsilon_0 \epsilon_{\mathrm{LRA}}\left(\mathbf{r}, \omega\right) \mathbf{E}\left(\mathbf{r}, \omega\right), \label{eq:LRA}
\end{equation}
in which $\epsilon_{\mathrm{LRA}}\left(\mathbf{r}, \omega\right)$ is the frequency and position dependant local dielectric function. This means that the material response at $\mathbf{r}$ depends solely on the driving field at $\mathbf{r}$. Although such a formalism can be used to study SPhPs in inhomogeneous systems \cite{Gubbin2016b}, it remains accurate only while dispersion of the material properties can be neglected. To move beyond this local theory we model the polar crystal as an isotropic continuum coupled to Maxwell's equations. Such an approach, detailed in Appendices \ref{APA} and \ref{APB}, leads to a theory describing a dispersive LO phonon branch existing at all the frequencies $\omega$ and (complex) wavevectors $\mathbf{k}$ satisfying {$\epsilon_{\mathrm{L}} \left(\omega, k \right) = 0$ \cite{Li1998}, where $\epsilon_{\mathrm{L}}$ is the longitudinal dielectric function
\begin{equation}
	\epsilon_{\mathrm{L}} \left(\omega, k\right) = \epsilon_{\infty} \frac{\omega_{\mathrm{L}}^2 - \omega\left(\omega + i \gamma \right) - \beta_{\mathrm{L}}^2 k^2}{\omega_{\mathrm{T}}^2 - \omega \left(\omega + i \gamma\right) - \beta_{\mathrm{L}}^2 k^2}.\label{eq:epslon}
\end{equation}
Transverse fields in the lattice are instead described by the transverse dielectric function 
\begin{equation}
	\epsilon_{\mathrm{T}} \left(\omega, k \right) = \epsilon_{\infty} \frac{\omega_{\mathrm{L}}^2 - \omega \left(\omega + i \gamma\right) - \beta_{\mathrm{T}}^2 k^2}{\omega_{\mathrm{T}}^2 - \omega \left(\omega + i \gamma\right) - \beta_{\mathrm{T}}^2 k^2}. \label{eq:epstra}
\end{equation}
and satisfy dispersion relation $\epsilon_{\mathrm{T}}\left(\omega, k \right) \omega^2 = c^2 k^2$.} In \Eqs{eq:epslon}{eq:epstra} $\beta_{\mathrm{L}}$ and  $\beta_{\mathrm{T}}$ are phenomenological velocities describing the LO and TO phonon dispersions respectively, {both of which are treated in a quadratic approximation \cite{TralleroGiner1992}.} {Both velocities may be extracted from \emph{ab initio} calculations or experimental measurements of the polar lattice phonon dispersion, by fitting a quadratic expansion to the low-wavevector region.} \\
Such an approach is analogous to established theories in nonlocal plasmonics, where a hydrodynamic description of the electron gas is typically employed  \cite{Fuchs1971,Ciraci2013,Schnitzer2016}, leading to smearing of electromagnetic hotspots, plasmon spill-in induced frequency shifts, and the emergence of confined longitudinal modes \cite{GarciadeAbajo2008, McMahon2009}. Notwithstanding the technical similarities, we find the impact of nonlocality on the optical response of phonon polaritons qualitatively differs from nonlocal plasmonics. This can be intuitively understood by considering Fig.~\ref{fig:Fig1}a, which illustrates longitudinal mode dispersion in plasmonic and phononic systems. In a metal the bulk plasma frequency blue shifts with increasing wavevector
\begin{equation}
	\omega = \sqrt{\omega_{\mathrm{P}}^2 + \beta^2 k^2}, \label{eq:Bpl}
\end{equation}
where $\beta$ is a characteristic velocity governing the dispersion of the plasma wave. This equation has real wavevector solutions for $\omega > \omega_{\mathrm{P}}$, illustrated by grey shading in Fig.~\ref{fig:Fig1}a. Surface plasmons, mediated by the negative dielectric function of the metal, are supported in the region $\omega_{\mathrm{SP}} < \omega_{\mathrm{P}}$, illustrated by the green shading. Here \Eq{eq:Bpl} only has evanescent solutions and longitudinal plasma waves decay with typical skin depth of order $1 \; \mathrm{\AA}$.\\
Optical phonons exhibit a red shift with increasing wavevector, as clear from Eq.~\ref{eq:epslon}. This means propagative longitudinal phonons coexist with the region where the dielectric function is negative, illustrated by the hatched region in Fig.~\ref{fig:Fig1}a. 
{Local theories imply delta-like screening charges at the particle surface. The charge profile is smoothed in a nonlocal theory by longitudinal modes excited at the particle edge which push charge away from the surface \cite{Ciraci2013}. In metallic particles this electronic spill-in causes a nonlocal change in electron density, leading to blue-shifted mode frequencies for $\omega < \omega_{\mathrm{P}}$ \cite{Raza2013}. Additional peaks appear for $\omega > \omega_{\mathrm{P}}$, where \Eq{eq:Bpl} has real wavevector solutions and the longitudinal waves can propagate into the sphere. Peaks correspond to quantised plasma waves in the sphere \cite{Raza2011, Rojas1988}. In the polar case we observe equivalent effective ionic charge spill-in, with the difference that the equation determining the longitudinal modes $\epsilon_{\mathrm{L}} (\omega, k ) = 0$ has real wavevector solutions for $\omega < \omega_{\mathrm{L}}$. The modes responsible for the spill-in are thus discrete propagative modes occupying the same spectral region as SPhPs.}\\
To illustrate the striking impact of this difference on the nonlocal optical features we show a comparison of extinction spectra calculated utilising nonlocal Mie theory for 10 nm diameter gold and 3C-SiC spheres in Fig.~\ref{fig:Fig1}b and Fig.~\ref{fig:Fig1}c respectively. The theory underlying this calculation is presented in the Appendices. \\

\subsection{Nonlocal Length Scales}
{Nonlocal systems can be assessed by a simple figure of merit, the skindepth of nonlocal excitations
\begin{equation}
	l = \frac{1}{\mathrm{Im} \left[ k \right]},
\end{equation}
where the wavevector is assumed perpendicular to the surface of the metal or polar crystal. To understand this we consider the LO phonon dispersion, given by the zeros of the longitudinal dielectric function in Eq.~\ref{eq:epslon}. 
The quality factor of optical phonon resonances is usually large enough to justify a lowest order expansion in the loss rate around $\gamma = 0$: $\gamma \ll \omega, \omega_{\mathrm{L}}$. We obtain
\begin{equation}
	\beta_{\mathrm{L}} k = \sqrt{\omega_{\mathrm{L}}^2 - \omega^2} - \frac{i \omega \gamma}{2 \sqrt{\omega_{\mathrm{L}}^2 - \omega^2}} + \mathcal{O}\left(\gamma^2\right).
	\label{betaLk}
\end{equation}
In the propagative phonon regime $\omega < \omega_{\mathrm{L}}$ the square-root yields a real quantity and the imaginary wavevector is given by the second term, leading to length scale
\begin{equation}
	l_{\mathrm{ph}} = \frac{2 \beta_{\mathrm{L}}  \sqrt{\omega_{\mathrm{L}}^2 - \omega^2}}{\omega \gamma}. \label{eq:phononlength}
\end{equation}
In the plasmonic case, using Eq.~\ref{eq:Bpl}, we obtain the equivalent of \Eq{betaLk}, but with imaginary and real parts inverted.  For $\omega < \omega_{\mathrm{P}}$ we find
\begin{equation}
	l_{\mathrm{pl}} = \frac{\beta}{\sqrt{\omega_{\mathrm{P}}^2 - \omega^2}}.
\end{equation}
Evaluating at the Fr{\"o}hlich frequency in each material we find $\mathrm{l_{ph}} \approx 12$ nm for 3C-SiC and $\mathrm{l_{pl}} \approx 1\; \mathrm{\AA}$ for gold.}

\section{Scattering from SiC nanospheres}
\begin{figure}
    \includegraphics[width=0.4\textwidth]{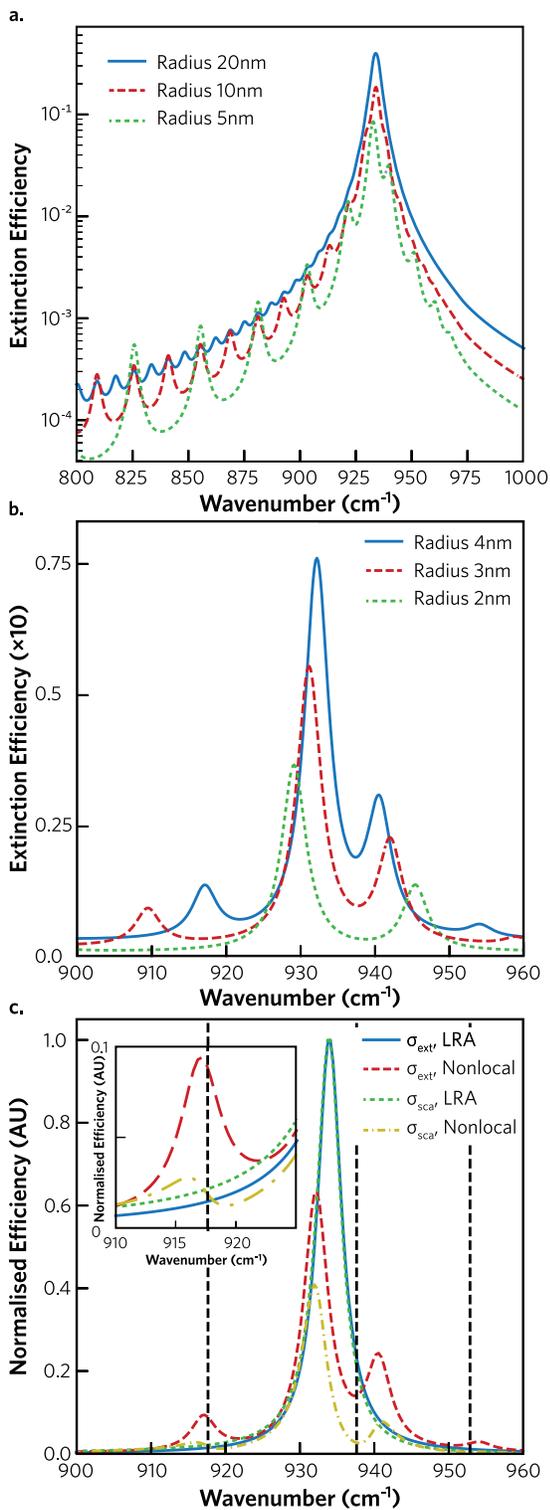}
	\caption{\label{fig:Fig2} Panels a) and b) show a comparison of nonlocal extinction efficiencies for 3C-SiC spheres of different radii. c) Detailed comparison of the extinction ($\sigma_{\mathrm{ext}}$) and scattering ($\sigma_{\mathrm{sca}}$) cross sections for a $4$ nm radius sphere in the local and nonlocal cases. Both curves have been normalised to unity so they may be presented on the same vertical scale. The inset is a zoom into the resonance near $905\;  \mathrm{cm}^{-1}$. The vertical dashed lines indicate the frequencies of the bare, discrete longitudinal resonances}
\end{figure}
In Fig.~\ref{fig:Fig2}a we plot the extinction cross section $\sigma_{\mathrm{ext}}$, for 3C-SiC spheres of $5, 10$ and $20$ nm radius, calculated using an extended Mie theory detailed in Appendix \ref{APC}.  In the local case deep-subwavelength spheres exhibit a dipolar Fr{\"o}hlich resonance where $\epsilon_{{\mathrm{LRA}}}\left(\omega_{\mathrm{F}}\right) = - 2$, at $\omega_{\mathrm{F}}\approx 934 \; \mathrm{cm}^{-1}$ for 3C-SiC. In nonlocal extinction spectra this resonance remains, however in smaller particles additional closely spaced peaks appear for $\omega < \omega_{\mathrm{L}}$. {These are the result of the screening charge induced at the particle surface. The induced charges repulse, producing density waves which spread from the interface into the nanoparticle bulk. As in this spectral region $\epsilon_{\mathrm{L}}\left(\omega, k\right) = 0$ has real wavevector solutions the excited longitudinal waves are propagative and have a discrete spectrum.} {These peaks become prominent when the particle radius reduces below $\approx 10$ nm, in agreement with the prediction of Eq.~\ref{eq:phononlength}.}  \\
 {
As the sphere radius decreases we observe level repulsion and the onset of strong coupling between the Fr{\"o}hlich resonance and the longitudinal modes. 
In a bulk crystal such coupling  is forbidden due to the orthogonal nature of longitudinal and transverse modes, which are only coupled by the system's interfaces. Accordingly, the effect becomes more visible in Fig.~\ref{fig:Fig2}b where we consider smaller radii and focus on the peak region. Note that this phenomenon is absent in the plasmonic case, being due to the existence of discrete propagative longitudinal modes which can become resonantly coupled to the Fr{\"o}hlich resonance. The coupling is proportional to the SPhP line-shape and is therefore enhanced for longitudinal modes near to resonance with the SPhP \cite{Giannini2011}.}\\
{When the SPhP is not on resonance with the LO modes coupling decreases and we instead recover behaviour typical of dark state hybridisation, in which a broad bright state hybridises with narrower dark states resulting in a Fano scattering spectra}\cite{Giannini2011}. Here the Fr{\"o}hlich resonance, radiative in the far field, plays the part of the bright state while longitudinal modes are dark. This behaviour is analogous to the weak phonon coupling previously observed in larger 4H-SiC particles, where the weak phonon frequency was determined by the lattice periodicity rather than particle morphology \cite{Gubbin2019, Ellis2016}. 
Fano interference acts to reduce the bright states far-field scattering, while increasing absorption as energy is funnelled into non-radiative dark states \cite{Gennaro2014,Simoncelli2018}. This is visible in the comparison of scattering and absorption efficiencies for a $4$ nm radius sphere in Fig.~\ref{fig:Fig2}c, where both cross sections are normalised to unity in order that they may be presented on the same vertical scale. On this plot vertical lines indicate the pure longitudinal resonances, calculated assuming the longitudinal electric field vanishes at the particle boundary. As shown in the inset, the longitudinal modes at $905$ and $950 \; \mathrm{cm}^{-1}$ suppress scattering while enhancing absorption, giving a Fano line profile. 

\subsection{Frequency Shifts}

\begin{figure}
	\includegraphics[width=0.4\textwidth]{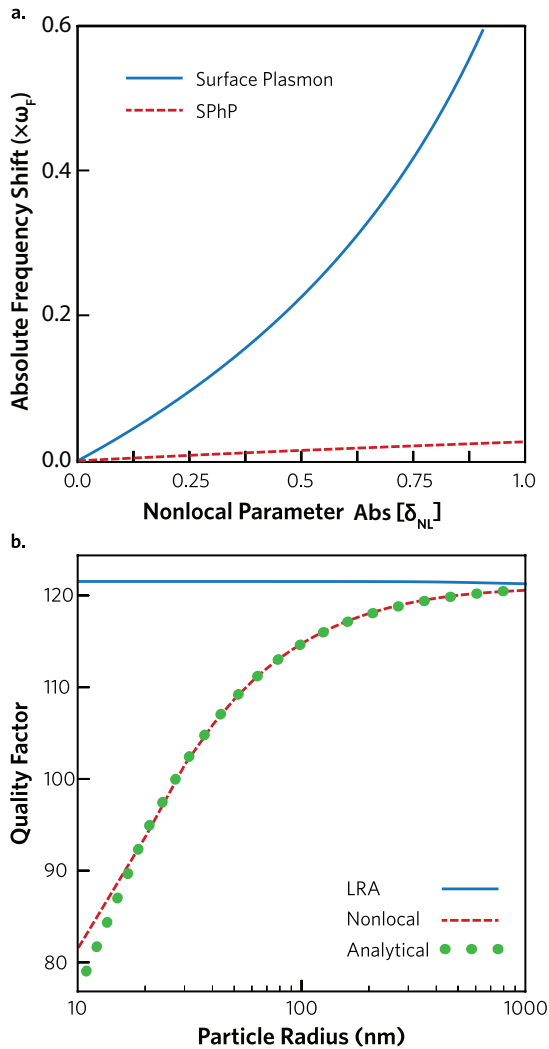}
	\caption{\label{fig:Fig3} a) Absolute nonlocal frequency shift of the main Fr{\"o}hlich peak in the extinction cross section as a function of the absolute value of the parameter quantifying the nonlocality. b) Local and nonlocal quality factors for the Fr\"ohlich resonance of a 3C-SiC sphere, the dots represent the analytical model in Eq.~\ref{eq:krei}}
\end{figure}

In plasmonic systems the dominant nonlocal effect is typically a blue shift in mode frequency \cite{Raza2011}, manifested in a quasi-static model as an increase in the effective local dielectric function. In the case of polar dielectrics the Fr{\"o}hlich resonance also undergoes a nonlocal shift, in this case to the red, but of much smaller amplitude. This can be verified from Fig.~\ref{fig:Fig3}a, in which we compare the absolute plasmonic and phononic frequency shifts from their local values $\omega_{\mathrm{F}}$ as a function of the absolute value of the dimensionless parameter quantifying nonlocal effects $\delta_{\mathrm{NL}}$, precisely defined in Appendix \ref{APC}. The reduced shift in the polar dielectric case is due to the narrow nature of the Reststrahlen band, which  leads to a faster frequency dispersion of the local dielectric function. The system can accommodate a change in the dielectric function due to a certain value of the nonlocal parameter $\delta_{\mathrm{NL}}$ by a smaller shift in frequency.

\subsection{Enhanced Nonlocal Broadening}

Another nonlocal effect often observed in the optics of small plasmonic particles is a size-dependent damping, initially described by Kreibig \cite{Kreibig1969}. This can be introduced phenomenologically in a hydrodynamic nonlocal theory as a diffusive term \cite{Mortensen2014,Ciraci2017}. Any broadening is size-dependent, requiring electrons to diffuse during the bare mode lifetime over a distance comparable to the field confinement length. The reason {why} size-dependent damping does not emerge from a pure hydrodynamic model in the absence of diffusion is because the longitudinal plasma waves excited in the electron gas obey \Eq{eq:Bpl}, and are thus evanescent. In the polar system studied in this paper the LO modes excited by the SPhPs are instead propagative, and can thus transport energy away from the interface. This allows them to act as an additional loss channel, leading to hydrodynamic Kreibig damping, which remains important even for relatively large particles.\\ 
{In order to calculate the size-dependent broadening we use a semi-analytical approximation of the full nonlocal Mie theory, described in Appendix C, which yields the frequency dependant polarisability of the sphere. SPhP mode-frequency and linewidth are then extracted by a Lorentzian fit.} In Fig.~\ref{fig:Fig3}b we plot the resulting quality factor of the fundamental Fr{\"o}hlich resonance for 3C-SiC spheres with radii between $10$ and $1000$ nm. In this region the particle is sufficiently large to support many closely spaced longitudinal modes, whose finite linewidth mean they form a continuum and do not appear as discrete peaks in the extinction cross section. The results show that for particles with radii less than $1\;\mu$m, coupling to these non-resonant modes leads to a broadening relative to the local result which grows with decreasing particle radius. We can form a model analogous to that of Kreibig to fit the data, assuming nonlocal damping rate of form
\begin{equation}
	\gamma_{\mathrm{NL}} = \gamma + \frac{\mathrm{A} \beta_{\mathrm{L}}}{\mathrm{R}}, \label{eq:krei}
\end{equation}
where $\gamma$ is the local damping rate and $\mathrm{R}$ is the nanosphere radius.\\
The dimensionless constant $\mathrm{A}$ describes the magnitude of the size-dependent damping. Fitting this formula to the numerical data we find $\mathrm{A} = 0.03$. {We can utilise this to calculate a typical length scale for which damping via LO phonon emission becomes dominant}
\begin{equation}
	r_{\mathrm{NL}} = \frac{\mathrm{A} \beta_{\mathrm{L}}}{\gamma} \approx 6.1 \; \mathrm{nm}, 
\end{equation}
which is comparable to that for diffusive damping in plasmonic systems where the characteristic velocity and bare damping rate are both enhanced by around 2 orders of magnitude \cite{Mortensen2014}.

\section{Nonlocality in planar nitride systems}

\subsection{Nonlocality in \texorpdfstring{A\MakeLowercase{l}N}{SiC} ENZ modes}

\begin{figure}
	\includegraphics[width=0.4\textwidth]{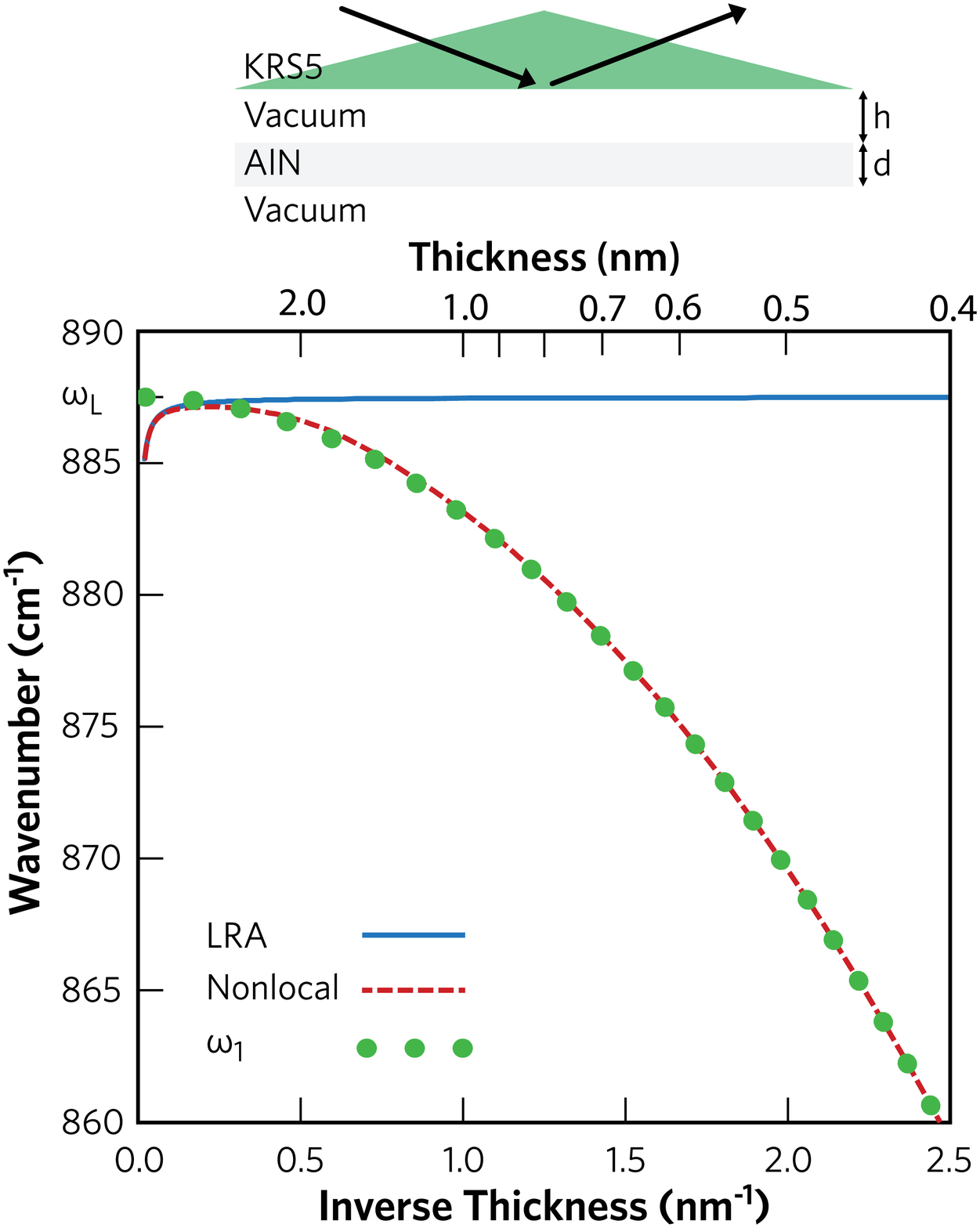}
	\caption{\label{fig:Fig4}  Comparison of the local (blue solid) and nonlocal (red dashed) ENZ mode frequencies in an AlN film as a function of inverse film thickness $d^{-1}$. The zone-centre longitudinal optical phonon frequency $\omega_{\mathrm{L}}$ is labelled on the axis and the $n = 1$ longitudinal mode energy from \Eq{eq:enzdomega} by green circles. Above the plot a sketch of the Otto configuration considered is shown.}
\end{figure}

Now we investigate nonlocal effects in freestanding ultrathin AlN films. Details of the approach used are provided in Appendix \ref{APD}. Optically thick films support degenerate SPhP modes at each interface. When the thickness $d$ decreases, eventually these hybridise yielding two distinct SPhP branches which are symmetric and antisymmetric superpositions of the modes on each interface. The symmetric mode is blue-shifted from the mode of the thick film and in the $d \to 0$ limit exists at the asymptotic zone-centre LO phonon frequency. As the dielectric function of the film vanishes at this frequency the mode is often referred to as an epsilon-near-zero (ENZ) mode \cite{Campione2015, Passler2018}. \\
Just like the SiC spheres studied in the previous section, a thin film also acts as a Fabry-Perot cavity for longitudinal phonons, supporting a series of discrete modes with quantised out-of-plane wavevector, satisfying $\eta = \frac{n \pi}{d}$, with $n$ a positive integer. For wavevectors in the film plane $k_{x} \ll \eta$ the LO phonon modes have frequencies
\begin{equation}
	\omega_n = \sqrt{\omega_{\mathrm{L}}^2 - \left(\frac{n \pi\beta_{\mathrm{L}} }{d}\right)^2}, \label{eq:enzdomega}
\end{equation}
a result analogous to that observed for ENZ modes in thin conductive films, where additional resonances obeying a similar equation were observed above the film plasma frequency \cite{deCeglia2018}. In a sufficiently thin sample the fundamental LO resonance $\omega_1$ can differ appreciably from the zone-center frequency $\omega_{\mathrm{L}}$, leading to a  shift in the upper edge of the Reststrahlen band. This is illustrated for an AlN film by the dots in Fig.~\ref{fig:Fig4}, which show the $n = 1$ solution of Eq.~\ref{eq:enzdomega} as a function of inverse film thickness. As the ENZ mode lies close to the LO phonon frequency, it can be expected to be extremely sensitive to such a shift.\\
To study this we calculate the frequency of the ENZ mode at constant in-plane wavevector $k_{\parallel} c = 1200 \; \mathrm{cm}^{-1}$ for an AlN film as a function of inverse film thickness in the local and nonlocal cases. In order to access the ENZ mode, which lies outside the light line, we utilise an Otto prism coupling configuration, schematically shown at the top of Fig.~\ref{fig:Fig4}, considering the AlN film of thickness $d$, surrounded by vacuum with a high refractive index ($n = 2.4$) KRS5 prism a distance $h$ above. Results shown in Fig.~\ref{fig:Fig4} demonstrate that nonlocal effects shift the ENZ resonance from its $\omega_{\mathrm{L}}$ frequency, having it to follow instead the effective edge of the Reststrahlen band $\omega_1$ for films less that $10$ nm thick. In the following we will see that similar shifts due to the ENZ confinement can be observed also by embedding the AlN between other materials, at frequencies for which the Reststrahlen 
bands do not overlap.

\subsection{Nonlocality in \texorpdfstring{A\MakeLowercase{l}N}{AlN}/\texorpdfstring{G\MakeLowercase{a}N}{GaN} atomic-scale superlattices}
\begin{figure}
    \includegraphics[width=0.4\textwidth]{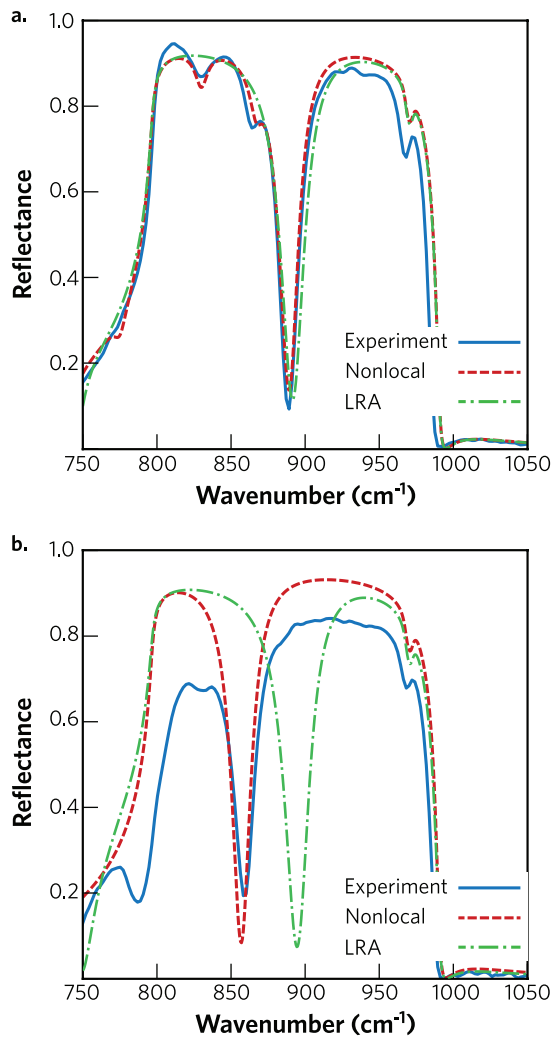}
	\caption{\label{fig:Fig5} Comparison of experimental reflectance data \cite{Ratchford2019} (blue solid), a local dielectric description (green dot-dashed) and a nonlocal description (red dashed) of nitride superlattices. Panel a) shows results for heterostructure A and b) for heterostructure B as described in the text.}
\end{figure}
 Recently the fabrication and optical characterisation of atomic-scale superlattices or crystal hybrids, comprised of alternating nanoscopic layers of GaN and AlN grown on a semi-insulating SiC substrate were reported \cite{Ratchford2019}. Such a work demonstrated the failure of a standard local response theory to describe the observed optical properties of the superlattice, requiring instead density functional perturbation theory to correctly reproduce its features  \cite{Paudel2009}.\\
Using a nonlocal extension of the scattering matrix approach we simulated the two heterostructures, labelled A and B, characterised in Ref. \cite{Ratchford2019}. Heterostructure A consists of 50 alternating layers of GaN ($2.2$ nm)/ AlN ($2.2$ nm) on a $50$ nm AlN buffer layer. Heterostructure B consists of 500 alternating layers of GaN ($1.4$ nm)/ AlN ($1.2$ nm). Both heterostructures are grown on a deep SiC substrate. In Fig.~\ref{fig:Fig5} we show experimental data, courtesy of the authors, with blue-solid lines and we compare them to our optimised nonlocal model (red dashed lines) and a separately optimised local theory (green dot-dashed lines). More details on the fitting procedure are presented in Appendix \ref{APE}.\\
Not only does the nonlocal theory provide a better fit, but we can recognise two sets of features which are correctly reproduced by the nonlocal theory while not captured at all by the local one. The first is photonic hybridisation with longitudinal modes, leading to visible peaks in heterostructure A around $835 \; \mathrm{cm}^{-1}$ and $865 \; \mathrm{cm}^{-1}$. These are analogous to the nonlocality-induced additional peaks visible in Fig.~\ref{fig:Fig2}. The second is the strong redshift of the resonance around $855\; \mathrm{cm}^{-1}$ for heterostructure B, which we already discussed in Fig.~\ref{fig:Fig4} for the freestanding film. {As previously noted, our nonlocal theory treats phonon dispersions as quadratic. This approximation reproduces well the phonon dispersion near the zone-centre phonon frequencies, but it diverges from the physical dispersion closer to the Brillouin zone edge \cite{Karch1994}. One consequence of the quadratic approximation is the failure of our model to correctly reproduce the dip observed at $795 \; \mathrm{cm}^{-1}$ in Fig.~\ref{fig:Fig5}b. The modes confined in the layers of the narrower structure have in fact larger wavevectors through \Eq{eq:enzdomega}, thus probing a region in which the quadratic dispersion underestimate the longitudinal phonon frequency, pushing the longitudinal modes at lower frequencies.
The eventual breakdown of the quadratic approximation is also why we constrain the plot region in Fig.~\ref{fig:Fig5} to around the SiC Reststrahlen region. It is {\it a priori} possible to overcome such a problem by considering wavevector-dependent, piecewise constant effective phonon velocities, able to reproduce the phonon dispersion over larger portions of the Brillouin zone, but the advantages of such an approach have to be weighted against the increase of required fitting parameters.}\\
{The calculated parameter for the AlN longitudinal velocity $\beta_{\mathrm{L}}^{\mathrm{AlN}} = 5.1 \times 10^5 \; \mathrm{cm\;s}^{-1}$ and damping $\gamma^{\mathrm{AlN}} = 10.3 \; \mathrm{cm}^{-1}$ leads through \Eq{eq:phononlength} to a predicted skin-depth of $\mathrm{l}_{\text{ph}} = 1.5 \; \mathrm{nm}$ at the $n = 1$ resonant frequency $855\; \mathrm{cm}^{-1}$. As for dielectric nanospheres, also in this case the use of \Eq{eq:phononlength} provides the right order of magnitude for the length scale at which nonlocal phenomena become important.}

\section{Discussion and Conclusions}
We studied the emergence of nonlocal effects in isotropic polar dielectric systems, demonstrating their importance for technologically relevant nanostructures.  We demonstrated that the negative dispersion of optical phonons, and the existence of a finite transverse frequency bounding the Reststrahlen band from below, lead to a rich and unique nonlocal phenomenology. In particular opening the possibility to resonantly couple photonic excitations to propagative longitudinal modes makes it possible to engineer hybrid longitudinal-transverse SPhPs, recently highlighted as an important stepping stone toward the realisation of electrically pumped SPhP devices \cite{Gubbin2019}, in generic nanoscale objects.\\
{ By modelling recently published experimental data, demonstrating that some of their unexplained features result directly from nonlocal effects, showing their immediate importance to the design of dielectric nanodevices, and providing a first experimental verification that our theory provides quantitatively correct results for systems down to few atomic lattices in size. 
This can mean that the physics at that level is already acceptably well described by a continuous theory or, alternatively, that the average effect of the microscopic degrees of freedom can be described by a renormalization of the effective nonlocal system parameters. Future experimental and numerical investigations will be required to gain a better understanding of the boundary of applicability of our approach. When used in conjunction with experimental data or microscopic calculations \cite{Chalopin2012, Paudel2009}, required to fix some of the model parameters, our theory thus allows for agile design and simulation of nanoscopic polar devices. Although in this paper all materials are treated as isotropic our model could be extended to consider anisotropic systems. This would entail numerical calculation of the modal dispersion, replacing the ionic equation of motion with it's anisotropic analogue.
We were able to implement our nonlocal theory in a number of numerical and analytical techniques adapted to different geometries (Mie theory, quasi-static, scattering matrix).  It remains an open question whether at least some of the nonlocal phenomenology due to the propagating longitudinal modes can be captured by using local models as in the plasmonic case \cite{Luo2013}, thus allowing for simple integration into existing commercial numerical software.}

\section*{Acknowledgements}
S.D.L. is a Royal Society Research Fellow. The authors acknowledge support from the Innovation Fund of the EPSRC Programme EP/M009122/1 and from the Royal Society grant RGF\textbackslash EA\textbackslash 181001.

\appendix

\section{Continuum dielectric model}
\label{APA}
We treat our polar crystal as an isotropic lattice, described in the continuum limit by \cite{Trallero-Giner1992, Li1998}
\begin{equation}
	\left[\omega_{\mathrm{T}}^2  - \omega\left(\omega + i \gamma\right) + \beta_{\mathrm{L}}^2 \nabla \left(\nabla \cdot \right) - \beta_{\mathrm{T}}^2 \nabla \times \nabla\times \right] \mathbf{X} =\frac{\mu}{\rho} \mathbf{E}, \label{eq:IonEOM}
\end{equation}
in which $\mathbf{X}$ is the relative ionic displacement, $\omega_{\mathrm{T}}$ is the transverse optical (TO) phonon frequency and $\rho$ and $\mu$ are the effective mass and charge densities. We suppress space (wavevector) and time (frequency) dependencies where possible. For simplicity we consider a single loss rate $\gamma$, independent of both frequency and polarization. The spatially dispersive terms in \Eq{eq:IonEOM} are calculable as the divergence of a tensor $\bar{\boldsymbol{\tau}}$
\begin{equation}
	\nabla \cdot \bar{\boldsymbol{\tau}} =  \nabla \cdot \left[ \beta_{\mathrm{T}}^2  \left( \nabla \mathbf{X} + \left(\nabla \mathbf{X}\right)^{\mathrm{T}}\right) + \left(\beta_{\mathrm{L}}^2 -  2 \beta_{\mathrm{T}}^2 \right) \nabla \cdot \mathbf{X}  \; \bar{\mathrm{I}} \right], \label{eq:Stress}
\end{equation}
in which $\bar{\mathrm{I}}$ is the identity tensor. In a model of acoustic phonons $\bar{\boldsymbol{\tau}}$ is the mechanical stress tensor, parameterised by the Lam{\'e} coefficients which describe the acoustic phonon velocities. As here we consider optical phonons the coefficients $\beta_{\mathrm{L}}, \beta_{\mathrm{T}}$ are instead phenomenological velocities which describe LO and TO phonon dispersions respectively \cite{TralleroGiner1992}. Note that this formalism only accounts for terms up to quadratic order in the phonon dispersion, potentially leading to deviation far from either phonon frequency. 
 {We can then introduce the total polarisation $\mathbf{P}$, which is a sum of the of the ionic polarisation $\mu \mathbf{X}$ and the high-frequency non-resonant background}
\begin{equation}
	\mathbf{P} = \mu \mathbf{X} + \epsilon_0\left(\epsilon_{\infty} - 1\right) \mathbf{E}, \label{eq:cons}
\end{equation}
in which $\epsilon_{\infty}$ is the high-frequency dielectric constant.\\
{The longitudinal and transverse dielectric functions, given in \Eq{eq:epslon} and \Eq{eq:epstra} of the main text, can then be derived. For the longitudinal mode $\nabla \times \mathbf{X} = 0$, and Fourier transforming \Eq{eq:IonEOM} we can write
\begin{equation}
	\mathbf{P}_{\mathrm{L}} = \left[ \epsilon_0 \left(\epsilon_{\infty} - 1 \right) +  \frac{\mu^2 / \rho}{\omega_{\mathrm{T}}^2 - \omega\left(\omega + i \gamma\right) - \beta_{\mathrm{L}}^2 k^2 }\right] \mathbf{E}_{\mathrm{L}}.
\end{equation}
Using the standard constitutive relation
\begin{equation}
	\mathbf{D} = \epsilon_0 \epsilon\left(\omega, k\right) \mathbf{E} = \epsilon_0 \mathbf{E} + \mathbf{P},
\end{equation}
we find
\begin{equation}
	\epsilon_{\mathrm{L}}\left(\omega, k \right) = \epsilon_{\infty} + \frac{\mu^2 / \rho \epsilon_0}{\omega_{\mathrm{T}}^2 - \omega\left(\omega + i \gamma\right) - \beta_{\mathrm{L}}^2 k^2 },
\end{equation}
which simplifies to that given in \Eq{eq:epslon} defining the LO phonon frequency  
\begin{equation}
\omega_{\mathrm{L}}^2=\omega_{\mathrm{T}}^2+\frac{\mu^2}{\rho\epsilon_0\epsilon_{\infty}}.
\end{equation}
A similar relation for the transverse dielectric function $\epsilon_{\mathrm{T}}\left(\omega, k\right)$ in \Eq{eq:epstra} can be derived considering instead $\nabla \cdot \mathbf{X} = 0$ in \Eq{eq:IonEOM}.}

\section{Additional boundary conditions}
\label{APB}

 In the LRA mode amplitudes at the interface between two piecewise continuous media are determined by application of the Maxwell boundary conditions (MBC) to electric and magnetic fields $\mathbf{E}$, and $\mathbf{H}$. As a nonlocal formalism entails consideration of additional fields in each layer, additional boundary conditions (ABC) are necessary to uniquely determine the model. 
In order to find the correct boundary conditions we use the ionic equation of motion \Eq{eq:IonEOM} to derive a hybrid Poynting vector describing the total energy transported in both the electromagnetic and mechanical fields. {We start from the Poynting's theorem
\begin{equation}
	\int_{\delta \Omega} \left(\mathbf{E} \times \mathbf{H} \right) \cdot \hat{\mathbf{n}} \,\mathrm{d S} 
	= - \int_{\Omega} \left[ \epsilon_0 \mathbf{E} \cdot \dot{\mathbf{E}} + \mathbf{E} \cdot \dot{\mathbf{P}} + \mu_0 \mathbf{H} \cdot \dot{\mathbf{H}}\right] \mathrm{d V}, \label{eq:Poyn}
\end{equation}
which relates the change in energy enclosed in volume $\mathrm{\Omega}$ with the flux through it's surface $\delta \Omega$. Substituting \Eq{eq:IonEOM} into \Eq{eq:Poyn} via the constitutive relation \Eq{eq:cons}, an approach proposed by Loudon \cite{Loudon1970}, we can write
\begin{align}
	&\int_{\delta \Omega} \left(\mathbf{E} \times \mathbf{H} \right) \cdot \hat{\mathbf{n}} \,\mathrm{d S} 
	= - \int_{\Omega} \biggr[ \epsilon_0 \epsilon_{\infty} \mathbf{E} \cdot \dot{\mathbf{E}} + \mu_0 \mathbf{H} \cdot \dot{\mathbf{H}} \nonumber \\
	& \quad \quad \quad + \rho \left(\omega_{\mathrm{T}}^2 - \omega \left(\omega + i \gamma\right) \right)\mathbf{X} \cdot \dot{\mathbf{X}}  + \rho \left(\nabla \cdot \bar{\boldsymbol{\tau}} \right) \cdot \dot{\mathbf{X}}
	\biggr] \mathrm{d V}.
\end{align}
The final term in this equation describes mechanical transport. It can be recast as
\begin{align}
	\int_{\Omega} \mathrm{d V} \rho \left[\nabla\cdot \bar{\boldsymbol{\tau}}_i \right] \dot{\mathrm{X}}_i &= \int_{\Omega} \mathrm{d V} \rho \left[ \nabla \cdot \left(\dot{\mathrm{X}}_i \bar{\boldsymbol{\tau}}_i\right) - \bar{\boldsymbol{\tau}}_i \cdot \nabla \dot{\mathrm{X}}_i \right] \nonumber \\
	&= \int_{\delta \Omega} \mathrm{d S} \rho \left[\dot{\mathrm{X}}_i \bar{\boldsymbol{\tau}}_i\right] \cdot \hat{\mathbf{n}} - \int_{\Omega} \mathrm{d V} \rho  \bar{\boldsymbol{\tau}}_i \cdot \nabla \dot{\mathrm{X}}_i,
\end{align}
where the latter term describes a nonlocal adjustment of the energy density and the former is an energy flux. Combining the flux term with the local Poynting flux from \Eq{eq:Poyn} we derive the hybrid Poynting vector
\begin{equation}
	\mathbf{S} = \mathbf{E} \times \mathbf{H} + \rho \bar{\boldsymbol{\tau} }\dot{\mathbf{X}},\label{eq:S}
\end{equation}
where the tensor $\bar{\boldsymbol{\tau}}$ is defined by \Eq{eq:Stress}.}
In order to ensure that interfaces act as neither energy sources or sinks, we impose continuity of the normal component of $\mathbf{S}$. This requires, in addition to the standard MBC, continuity of $\mathbf{X}$ and of the normal components of the tensor $\bar{\boldsymbol{\tau}}$.\\
At an interface between nonlocal and local media the boundary conditions are under-specified. This is a problem  well known in the case of exciton polaritons, where multiple solutions have been proposed, derived either from microscopic considerations \cite{Pekar1959}, by application of symmetry conditions to the fields \cite{Fuchs1971} {or by considering explicitly the nonlocal surface potential near the boundary \cite{Ruppin1984}}. The most commonly utilised ABC are the Pekar-Ridley \cite{Pekar1959} and Fuchs-Kliewer \cite{Fuchs1971} which, assuming for definiteness an interface lying in the $xy$ plane read $\mathbf{X} = 0$ and $\partial_z \mathrm{X}_{\{x,y\}} = \mathrm{X}_z = 0$ respectively. Both of these choices satisfy $\rho \bar{\boldsymbol{\tau}} \dot{\mathbf{X}} \cdot \hat{\mathbf{z}} = 0$, ensuring conservation of energy across the interface when the mechanical excitations cannot propagate ({\it e.g.}, at a dielectric-vacuum interface). \\
In the isotropic model we are considering the additional high-momentum transverse mode coupled by the nonlocality is closely resonant with the bare TO phonon, and thus it has a negligible associated electric field. ABC then predominantly mix the low-momentum transverse photon-like mode and the dispersive LO excitation. The condition on $\mathrm{X}_{\{x,y\}}$ is then trivially satisfied without affecting the optical response. This means that calculations of observables such as reflectance utilising the Fuchs-Kliewer and Pekar-Ridley boundary conditions are practically indistinguishable. We then use Fuchs-Kliewer type ABC as they are derived by symmetry arguments, and they are  equivalent to assuming specular reflection from the interface, which is the relevant symmetry in our case.

\section{Scattering from dielectric spheres}
\label{APC}

In order to include nonlocal effects in the treatment of the optical response of small dielectric spheres we used an extended Mie theory, conceptually similar to previous nonlocal extensions of Mie theory utilised to describe nonlocal effects in noble metallic systems \cite{Raza2011} and described fully in the Supplementary information. In order to test our calculations, and to provide a better understanding of the results, we also used a semi-analytic model valid in the quasi-static limit, that is assuming the impinging radiation field does not vary appreciably over the sphere diameter. \cite{Raza2013}. This is a good approximation for the parameters considered in this paper, with radii $\mathrm{R}$ normalised over the resonant wavelength of the order of $0.01$. For the parameters presented in Fig.~\ref{fig:Fig2}b the extinction cross section given in \Eq{eq:quasi} is indistinguishable from the full Mie theory.\\
In the quasi-static limit the polarisability of a sphere in vacuum has the form 
\begin{align}
	\alpha = 4 \pi \mathrm{R}^3 \frac{\tilde{\epsilon}\left(\omega\right) -  1}{\tilde{\epsilon}\left(\omega\right) + 2},
	\label{eq:alpha}
\end{align}
where  
\begin{equation}
	\tilde{\epsilon}\left(\omega\right) = \frac{\epsilon_{\mathrm{LRA}}\left(\omega\right)}{1 + \delta_{\mathrm{NL}}}, \label{eq:qnls}
\end{equation}
is the local dielectric function renormalised by nonlocal effects through the nonlocal parameter
\begin{equation}
	\delta_{\mathrm{NL}} =  \frac{1}{\xi \mathrm{R}} \frac{\epsilon_{\mathrm{LRA}}\left(\omega\right) - \epsilon_{\infty}}{\epsilon_{\infty}} \frac{j_1 \left(\xi \mathrm{R}\right)}{j_1'\left(\xi \mathrm{R}\right)},
\end{equation}
with $j_1(x)$ and $j_1'(x)$ a spherical Bessel function of the first kind and its derivative, and $\xi$ is the longitudinal mode wavevector defined by the equation
\begin{align}
\xi^2 = \left[\omega_{\mathrm{L}}^2 - \omega \left(\omega + i \gamma\right) \right] / \beta_{\mathrm{L}}^2. \label{eq:kL} 
\end{align}
Within this formalism we can quickly evaluate the extinction cross section
\begin{equation}
	\sigma_{\mathrm{ext}} = \frac{1}{\pi \mathrm{R}^2} \frac{\omega}{c} \left[\frac{\omega^3}{6 \pi c^3} \lvert \alpha \rvert^2 +  \mathrm{Im}\left(\alpha\right) \right].\label{eq:quasi}
\end{equation} 
Further details on both the extended Mie theory and on the semi-analytic quasi-static theory can be found in Appendix \ref{APC2}.\\
{
In the 3C-SiC case we use transverse optic phonon frequency $\omega_{\mathrm{T}} = 796.1 \; \mathrm{cm}^{-1}$, longitudinal optic phonon frequency $\omega_{\mathrm{L}} =  973 \; \mathrm{cm}^{-1}$ \cite{Caldwell2015a}, high-frequency dielectric constant $\epsilon_{\infty} = 6.52$, damping rate $\gamma = 4\; \mathrm{cm}^{-1}$, and nonlocal velocities $\beta_{\mathrm{T}}^{SiC} = 9.15 \times 10^5 \; \mathrm{cm\;s}^{-1}, \; \beta_{\mathrm{L}}^{SiC} = 15.39 \times 10^5 \; \mathrm{cm\; s}^{-1}$ approximated from studies of phonon dispersion \cite{Karch1994}. For the LO phonon damping rate we extrapolate the low temperature results of Debarnardi {\it et al.} to room temperature, assuming a 2 phonon decay channel \cite{Debernardi1999}.\\
In the gold case we use plasma frequency $\omega_{\mathrm{p}} =  72670 \; \mathrm{cm}^{-1}$, damping rate $\gamma = 580 \; \mathrm{cm}^{-1}$ \cite{Derkachova2015} and nonlocal velocity $\beta_{\mathrm{p}} = 1.08 \times 10^{8} \; \mathrm{cm \; s}^{-1}$ \cite{Luo2013, Ciraci2012}.}

\section{Nonlocal Mie Theory}
\label{APC2}
{
In Appendix \ref{APC} we derived the extinction cross section in the quasistatic limit for a nanosphere supporting longitudinal nonlocal modes. To consider transverse nonlocality it is necessary to develop a full nonlocal Mie theory. When considering the mechanical boundary conditions it is necessary to construct a set of spherical vector harmonics which also yield spherical vector harmonics under application of the divergence and curl operators. A suitable choice are \cite{Barrera1985}
\begin{align}
	\mathbf{Y}_{p l m}\left(\theta, \phi\right) &= \hat{\mathbf{e}}_r \mathrm{Y}_{p l m}\left(\theta, \phi\right),\\
	\boldsymbol{\Psi}_{p l m}\left(\theta, \phi\right) &= r \nabla \mathrm{Y}_{p l m}\left(\theta, \phi\right),\\
	\boldsymbol{\Phi}_{p l m} \left(\theta, \phi\right) &= \mathbf{r} \times \nabla \mathrm{Y}_{p l m}\left(\theta, \phi\right),
\end{align}
where $\mathrm{Y}_{p l m}(\theta,\phi)$ are scalar spherical harmonics with $p=\{e,o\}$ denoting even or odd parity in $\phi$  
\begin{align}
	\mathrm{Y}_{e l m} &= \mathrm{P}_{lm}\left(\cos \theta\right) \cos m \phi, \\ \mathrm{Y}_{o l m} &= \mathrm{P}_{lm}\left(\cos \theta\right) \sin m \phi,
\end{align}
and $\mathrm{P}_{lm}\left(\cos \theta\right)$ are the associated Legendre polynomials. Single-valued vector fields vanishing at infinity ($\mathbf{F}$) can be written in this basis
\begin{equation}
	\mathbf{F} = \sum_{p l m} \left[ \mathrm{V}_{p l m} \left(r\right) \mathbf{Y}_{p l m} +  \mathrm{V}_{plm}^{(1)}\left(r\right) \boldsymbol{\Psi}_{plm} +  \mathrm{V}_{plm}^{(2)}\left(r\right) \boldsymbol{\Phi}_{plm}\right], \label{eq:field_dec}
\end{equation}
where expansion coefficients can be found exploiting the orthogonality of the spherical vector harmonics. For example 
\begin{equation}
	\mathrm{V}_{plm}\left(r\right) = \frac{\int \mathrm{d}^3r\, \mathbf{F} \cdot \mathbf{Y}_{plm} }{\int \mathrm{d}^3r\,  \mathbf{Y}_{plm} \cdot \mathbf{Y}_{plm}}.
\end{equation}
This choice of spherical harmonics allows for segregation into longitudinal and transverse modes. Modes proportional to $\boldsymbol{\Phi}_{plm}$ are always transverse while those proportional to $\mathbf{Y}_{plm}, \boldsymbol{\Psi}_{plm}$ can be decomposed into transverse and longitudinal components satisfying respectively the equations
\begin{align}
	\frac{2  \mathrm{V}_{plm } \left(r\right)}{r}  +  \frac{d}{dr}\mathrm{V}_{plm} \left(r\right) &= \frac{l \left(l + 1\right)}{r}  \mathrm{V}_{plm}^{(1)} \left(r\right),\label{eq:Tcond} \\
	\frac{\mathrm{V}_{plm}^{(1)}\left(r\right)}{r} + \frac{d}{dr}\mathrm{V}_{plm}^{(1)}\left(r\right)&=\frac{\mathrm{V}_{plm}\left(r\right)}{r}.\label{eq:Lcond}
\end{align}
This theory can be applied to any layered spherical resonator supporting longitudinal and transverse nonlocality. As an example we calculate the fields in a single layer nanosphere of radius $R$, supporting a longitudinal mode. An impinging plane wave can be expanded in the region $r > R$ as
\begin{align}
	\mathbf{E}_\mathrm{I} &=  \mathrm{E}_0 e^{i k z} \hat{\mathbf{x}}\\
	&= 
	 \mathrm{E}_0 \sum_{l = 1}^{\infty} i^l \biggr[ \frac{l \left(l + 1\right) j_l\left(\rho \right)}{\rho} \mathbf{Y}_{el1}   
	 + \left( \frac{j_l\left(\rho\right)}{\rho} + j_l'\left(\rho\right)\right) \boldsymbol{\Psi}_{el1} \nonumber\\
	 &\quad \quad  \quad \quad \quad \quad  - j_l\left(\rho\right)\boldsymbol{\Phi}_{ol1}\biggr],\nonumber
\end{align}
where $\mathrm{E}_0$ is the field intensity and $k$ the wavenumber outside the sphere, $j_l(\rho)$ is a spherical Bessel function of the first kind and $\rho = k r$. The scattered field in region $r > R$ can be similarly written
\begin{align}
	\mathbf{E}_\mathrm{S} = -\mathrm{E}_0 \sum_{l = 1}^{\infty} i^l \biggr[ a_l \biggr(&\frac{l \left(l + 1\right) h_l^{(1)}\left(\rho \right)}{\rho} \mathbf{Y}_{el1} \nonumber \\
	& + \frac{\left[ \rho h_l^{(1)}\left(\rho\right)\right]'}{\rho}  \boldsymbol{\Psi}_{el1}\biggr) - b_l h_l^{(1)}\left(\rho\right)\boldsymbol{\Phi}_{ol1}\biggr],
\end{align}
where we enforced transversality through \Eq{eq:Tcond}, $a_l, b_l$ are the Mie scattering coefficients and $h_l^{(1)}(\rho)$ is a spherical Hankel function of the first type, chosen to ensure an outgoing spherical wave in the limit $\rho \to \infty$.\\
The transverse field in the region $r < R$ can be written
\begin{align}
	\mathbf{E}_\mathrm{T} = \mathrm{E}_0 \sum_{l = 1}^{\infty} i^l \biggr[ c_l \biggr[& \frac{l \left(l + 1\right) j_l\left(\rho_T \right)}{\rho_T} \mathbf{Y}_{el1}
	 \nonumber \\
	 &+ \frac{\left[ \rho_t j_l\left(\rho_T\right)\right]'}{\rho_T}  \boldsymbol{\Psi}_{el1}\biggr] - d_l j_l\left(\rho_T\right)\boldsymbol{\Phi}_{ol1}\biggr],
\end{align}
where $\rho_T = \sqrt{\epsilon \left(\omega\right)} k_0 r$ and $c_l, d_l$ are unknown coefficients. The longitudinal phonon electric field inside the particle is given by
\begin{align}
	\mathbf{E}_\mathrm{L} &= \mathrm{E}_0 \sum_{l = 1}^{\infty} i^l  g_l \left[\frac{\left[ \rho_L j_l\left(\rho_L \right)\right]'}{\rho_L} \mathbf{Y}_{el1} +  \frac{j_l\left(\rho_L\right)}{\rho_L}  \boldsymbol{\Psi}_{el1}\right], \label{eq:SphEL}
\end{align}
where we used \Eq{eq:Lcond} and $\rho_L = \xi r$, where $\xi$ is the longitudinal wavevector defined in \Eq{eq:kL}.\\
The unknown coefficients $a_l, b_l, c_l, d_l, g_l$ can be calculated by application of the appropriate boundary conditions. In this case these are the standard Maxwell boundary conditions on the components of $\mathbf{E}$ and $\mathbf{H}$ parallel to the sphere surface, in addition to the boundary condition on the radial component of the ionic displacement $\mathbf{X}$. In conjunction with continuity of the radial displacement field $\mathbf{D}$ this boundary condition can be recast as one on the radial component of $\epsilon_{\infty} \mathbf{E}$. This leads to extinction and scattering cross sections
\begin{align}
	\sigma_{\mathrm{ext}} &= \frac{2 \pi}{k^2} \sum_{l = 1}^{\infty} \left( 2 l + 1\right) \mathrm{Re}\left[a_l + b_l\right],\\
	\sigma_{\mathrm{sca}} &= \frac{2 \pi}{k^2} \sum_{l = 1}^{\infty} \left( 2 l + 1\right) \left(\lvert a_l \rvert^2 + \lvert b_l \rvert^2\right).
\end{align}}

\section{Nonlocal model of a layered dielectric}
\label{APD}
{The application of the nonlocal theory to a layered structure can be performed by calculating a complete set of modes for each layer and then applying MBC and ABC to match their amplitudes at the boundaries. In the case of the four layers considered for the AlN-vacuum ENZ resonance (prism--vacuum--AlN--vacuum) this leads to $12$ independent modes: two transverse counter-propagating modes for each non-active layer, and four transverse and two longitudinal modes for the AlN slab. The AlN slab (region 2) occupies $-d<z<0$, vacuum occupies regions 1 $z<-d$ and 3 $h> z>0$, and region 4 $z>h$ contains a high-index prism. The photon fields in the multilayer stack are given by
\begin{align} 
	\mathbf{E}_{1} &= \mathrm{B}_1 \frac{\sqrt{\epsilon_1} k_0}{\alpha_1} \hat{\mathbf{p}}_{+ 1} e^{\alpha_1 \left(z + d\right)} e^{i k_x x} e^{i \omega t},\\
	\mathbf{E}_{3} &= \frac{\sqrt{\epsilon_3} k_0}{\alpha_3} \left[ \mathrm{B}_{3}^{\mathrm{a}} \hat{\mathbf{p}}_{+ 3}   e^{  \alpha_3 (z - h)} + \mathrm{B}_{3}^{\mathrm{b}} \hat{\mathbf{p}}_{- 3} e^{  - \alpha_3 (z- h)}  \right] e^{i k_x x} e^{i \omega t},\\
	\mathbf{E}_{4} &= \frac{\sqrt{\epsilon_4} k_0}{\alpha_4} \left[ \mathrm{B}_{4}^{\mathrm{a}} \hat{\mathbf{p}}_{+ 4} e^{ \alpha_4 (z - h)}  + \mathrm{B}_{4}^{\mathrm{b}} \hat{\mathbf{p}}_{- 4}  e^{  - \alpha_4 (z - h)}  \right] e^{i k_x x} e^{i \omega t},
\end{align}
in which $k_x$ is the in-plane wavevector, $\omega$ the mode frequency, $\mathrm{B}_i$ are unknown coefficients to be determined by the boundary conditions and $\epsilon_i$ is the $i$th layers dispersionless dielectric function. The unit vector is given by
\begin{equation}
	\hat{\mathbf{p}}_{\pm i} \left(\alpha\right) = \frac{\alpha_i \hat{\mathbf{x}} \mp i k_x}{\sqrt{\epsilon_i} k_0},
\end{equation}
and the out-of-plane transverse wavevector in layer $i = 1, 3, 4$ is
\begin{equation}
	\alpha_i = \sqrt{k_x^2 - \epsilon_i \omega^2}.
\end{equation}
Assuming illumination from medium $4$ we can set $\mathrm{B}_{4}^{\mathrm{a}} = 1$ and recognise $\mathrm{B}_{4}^{\mathrm{b}}$ as the reflectance.\\
In the nonlocal AlN film two transverse and one longitudinal mode are supported. The electric field profile for TO phonons and photons can be written
\begin{align}
	\mathbf{E}_{2, \mathrm{T}} = \biggr[& \mathrm{B}_{\pm}^{\mathrm{a}} \hat{\mathbf{p}}_{+ 2} \left(\hat{\mathbf{x}} - \frac{i k_x}{\alpha_{\pm}} \hat{\mathbf{z}}\right)   e^{  \alpha_\pm (z + d )} \nonumber \\
	&+ \mathrm{B}_{\pm}^{\mathrm{b}} \left(\hat{\mathbf{x}} + \frac{i k_x}{\alpha_{\pm}} \hat{\mathbf{z}}\right) e^{  - \alpha_\pm (z + d)}  \biggr] e^{i k_x x} e^{i \omega t},
\end{align}
where $\alpha_{\pm}$ represents the two solutions to the nonlocal dispersion relation $\epsilon_{\mathrm{T}} \left(\omega, k\right) \omega^2 = c^2 k^2$. The longitudinal mode field can be written
\begin{align} 
\mathbf{E}_{\mathrm{L}}^{2} = \biggr[ &\mathrm{A}^{\mathrm{a}} \left( \hat{\mathbf{x}} - \frac{i \eta}{k_x} \hat{\mathbf{z}}\right) e^{ \eta \left(z + d\right)}  \nonumber \\
	& + \mathrm{A}^{\mathrm{b}} \left( \hat{\mathbf{x}} + \frac{i \eta}{k_x} \hat{\mathbf{z}}\right) e^{ - \eta \left(z + d\right)}\biggr] e^{i k_x x} e^{i \omega t},
\end{align}
where  $\mathrm{A}^{\{\mathrm{a,b}\}}$ are further undetermined coefficients and the out-of-plane wavevector
\begin{equation} 
	\eta = \sqrt{k_x^2 - \frac{\omega_{\mathrm{L}}^2 - \omega \left(\omega + i \gamma\right)}{\beta_{\mathrm{L}}^2}},\label{eq:lwv} 
\end{equation}
is the solution to $\epsilon_{L}\left(\omega, k \right) = 0$.\\
From these electric fields the fields entering the boundary conditions are calculable. The magnetic field is calculable by Maxwell-Faraday equation. As additional boundary conditions we constrain the in- and out-of-plane components of the ionic displacement $\mathbf{X}$. The AlN film is clad in vacuum where $\mathbf{X} = 0$. As the nonlocal model eliminates charge surface charging the normal component of the displacement field
\begin{equation}
	\mathrm{D}_z = \epsilon_0 \epsilon_{\infty} \mathrm{E}_z + \mu \mathrm{X}_z
\end{equation}
 is continuous at the AlN boundary. The additional boundary condition on the perpendicular component of $\mathbf{X} = 0$ can therefore be recast to one on $\epsilon_{\infty} \mathrm{E}_z$. The final boundary condition is $\mathrm{X}_{\parallel} = 0$. Applying these conditions yields a $10 \times 10$ matrix problem for the unknown coefficients $\mathrm{A, B}$.}

\section{Fitting procedure}
\label{APE}
Transfer matrix algorithms, the tool of choice to model optical properties of planar semiconducting heterostructures, are generally unstable in the limit of thick absorbing layers. As in a nonlocal model the phonon mode out-of-plane wavevectors are very large this problem emerges even when considering optically thin films. This instability forced us to utilise a scattering matrix approach. In a scattering matrix model we start by calculating the matrix of the $p^{\text{th}}$ layer $\mathrm{s}^{(p)}$ which relates the 
up- and down-propagating incoming modes $\mathbf{u}^{(p)}$ and $\mathbf{d}^{(p+1)}$ to the outgoing ones  
$\mathbf{u}^{(p+1)}$ and $\mathbf{d}^{(p)}$.
The scattering matrix for the full $P$-layer stack,  
\begin{eqnarray}
	 \left[ \begin{array}{c}
 	\mathbf{u}^{(P+1)}\\
 	\mathbf{d}^{(0)} 
 \end{array}\right] = {\mathrm{S}}^{(P)} \left[ \begin{array}{c}
 	\mathbf{u}^{(0)} \\
 	\mathbf{d}^{(P+1)} 
 \end{array}\right],\\\nonumber
\end{eqnarray}
can then be obtained by calculating the Redheffer star product of the single-layer matrices.
We then fitted the experimental data using both local and the nonlocal theories. In both cases we allowed the layers' thickness to vary around the nominal value in order to take into account atomic intercalation or screening \cite{Chalopin2012} at the interface, which reduces the effective thickness which can be described by bulk crystal structure. Consistently with such a picture we found optimal thickness reductions roughly equal to one lattice constant per interface. \\
The longitudinal phonon velocity in the non-resonant GaN layer was fixed at $\beta_{\mathrm{L}}^{GaN}=6.5\times10^5 \; \mathrm{cm \; s}^{-1}$, approximated from numerical phonon dispersion curves \cite{Bungaro2000}. In order to reduce the fitting parameters we neglected TO phonon velocities, which have a very limited impact on the calculated spectra, thus imposing $\beta_{\mathrm{T}}^{AlN}=\beta_{\mathrm{T}}^{GaN}=0$ also in the nonlocal theory. Our fitting procedure used two unknown parameters for the local theory and three in the nonlocal one: the shift in layer thickness $\delta$, the damping rate in the AlN $\gamma^{AlN}$, and the longitudinal velocity in the AlN layer $\beta_{\mathrm{L}}^{AlN}$, the latter being  unique to the nonlocal theory. Optimal parameters are found to be $\beta_{\mathrm{L}}^{AlN}= 5.1 \times 10^5 \; \mathrm{cm \; s}^{-1}$, $\delta = 0.94 \; \mathrm{nm}$ and $\gamma^{AlN} = 10.3 \; \mathrm{cm}^{-1}$.

\bibliographystyle{naturemag}
\bibliography{Bibliography}

\end{document}